\documentstyle[amsmath,amssymb,epsf,mncite]{mn}
\makeatletter\ifSFB@referee\fi\makeatother

\pubyear{0000}
\volume{000}
\pagerange{1--\pageref{lastpage}}
\def\kms{\ifmmode{{\rm km}\,{\rm s}^{-1}}\else{km\,s$^-1$}\fi}

\begin{document}

\date{Version of \today.}

\title[CMB polarization and ionization history]{
Cosmic Microwave Background polarization and the ionization history of
the Universe}

\newcounter{fusspilz}\setcounter{fusspilz}{0}
\def\fusspilz{{\stepcounter{fusspilz}\fnsymbol{fusspilz}}}
\author[Pavel Naselsky {\em et al.}]{
Pavel Naselsky$^{1,\fusspilz}$,
Jens Schmalzing$^{1,2,\fusspilz}$, \cr
Jesper Sommer-Larsen$^{1,\fusspilz}$, and
Steen Hannested$^{3,\fusspilz}$
\setcounter{fusspilz}{0}
\vspace*{1ex}\\
$^1$
Teoretisk Astrofysik Center,
Juliane Maries Vej 30,
DK-2100 K\o benhavn \O,
Denmark.
\\
$^2$
Ludwig--Maximilians--Universit\"at,
Theresienstra{\ss}e 37,
80333 M\"unchen, Germany.
\\
$^3$
NORDITA,
Blegdamsvej 17,
DK-2100 K\o benhavn \O,
Denmark.
\\
$^\fusspilz$email naselsky@tac.dk \\
$^\fusspilz$email jens@tac.dk \\
$^\fusspilz$email jslarsen@tac.dk \\
$^\fusspilz$email steen@nordita.dk \\
}

\maketitle

\begin{abstract}
We point out that polarization measurements as planned for the
upcoming PLANCK mission can significantly enhance the accuracy of
cosmic parameter estimation compared to the temperature anisotropy
spectrum alone.  In order to illustrate this, we consider a standard
cosmological model and several modifications that adjust one parameter
each to fit the recently published Maxima-1 data.  While all models
produce acceptable fits as far as the power spectrum is concerned,
their corresponding polarization spectra differ widely.  The strongest
differences are expected for a model with delayed recombination,
reflecting the fact that polarization measurements are most sensitive
to the processes governing the epoch of recombination.
\end{abstract}


Observations of the Cosmic Microwave Background (CMB) are fundamental
for our understanding of one of the most important epochs in the early
history of the Universe.  After a number of successful experiments,
among them COBE {\cite{bennett1996}}, BOOMERANG
{\cite{bernardis2000}}, Maxima-1 {\cite{hanany2000}}, CBI
{\cite{padin2001}} to mention just a few, much attention is now
focused on the estimation of cosmological parameters such as the
densities of baryons $\Omega_b$ and dark matter $\Omega_m$ and
cosmological constant $\Omega_{\Lambda}$.  In addition, preliminar
information about the Hubble constant $H_0=100h$, the power spectrum
of initial adiabatic perturbation, and the ionisation history of the
Universe are available.

Fitting the CMB anisotropy power spectrum to the observational
BOOMERANG and Maxima-1 data confirms the prediction of the standard
inflation scenario that the Universe is flat, i.e. that the total
amount of matter and vacuum energy has the critical value -- to an
accuracy of $10\%$, $\Omega_b+\Omega_m+\Omega_\Lambda=1$.  However, as
was already pointed out by, e.g.,
{\scite{tegmark2000:new,white2000,lesgourgues2000}}, the data
furthermore yield a baryon fraction $\Omega_bh^{2}$ significantly
larger than the value expected from standard Big Bang nucleosynthesis
(SBBN) supported by $He^4$ and deuterium mass fraction measurements.
These discrepancies stimulate discussion about possible alternatives
that either alter SBBN, e.g. by non-vanishing neutrino chemical
potentials {\cite{esposito2001,mangano2000}}, or add additional
non-standard parameters to the cosmological model, such as a tilt of
the power spectrum {\cite{tegmark2000:current}} or secondary
ionization of hydrogen
{\cite{tegmark2000:new,schmalzing2000:constraints}}.  All these
factors should be taken into account in the reconstruction of the
history of the cosmological expansion, and therefore in the
development of methods for cosmological parameter estimation for
future high precision experiments like MAP and PLANCK.

So far, however, little attention has been paid to a crucial part of
the modelling of the CMB anisotropy spectrum -- the history of
hydrogen and helium recombination at redshift $z\sim10^3$.  The
standard model of hydrogen recombination in the baryonic Universe was
suggested by {\cite{peebles1968}} and {\cite{zeldovich1968}} and
generalized to the dark matter models by {\cite{zabotin1982}} and
{\cite{jones1985}}.  A more precise model of recombination taking into
account $He^4$ was developed by {\cite{seager1999}}.  Recently, in
order to test the sensitivity of methods and experiments to the
detailed ionization history around recombination, this standard model
of recombination has been modified in various ways
{\cite{avelino2000,battye2001,hannested2001,landau2000}}.

{\cite{peebles2000}} discuss some modification of the primordial
ionization history of the hydrogen plasma at $z\sim10^3$ in the
framework of a delayed recombination model, that could arise in
certain scenarios with a decaying heavy particle
{\cite{sarkar1983,scott1991,ellis1992,doroshkevich2001}}, primordial
black holes evaporation {\cite{naselsky1978,naselsky1987}}, or cosmic
string wakes {\cite{weller1999}}.  In addition to the 10 cosmological
parameters discussed by {\cite{tegmark2000:current}} in connection
with the CMB anisotropy power spectrum, the possible delay of
recombination introduces another one, namely $\varepsilon_\alpha$,
related to the production of additional Lyman-$\alpha$ resonance
photons in the background of primordial radiation.  Their fraction
$n_{\alpha}$ is determined by the simple equation
$\frac{\mathrm{d}n_{\alpha}}{dt} = \varepsilon_{\alpha}H(t)n_H$, where
$H(t)$ is the Hubble parameter and $n_H$ is the neutral hydrogen
concentration.  The influence of such additional Lyman-$\alpha$
photons on the rate of ionization is fairly straightforward.  In
comparison to standard recombination, they compensate for the decrease
in the number of resonance quanta due to cosmological expansion and
lead to an increase of the fraction of ionization, which becomes
larger with the $\varepsilon_\alpha$ parameter.  This process
increases the dissipation length for initial adiabatic perturbations
and therefore suppresses the secondary Doppler-peaks.  It also shifts
the anisotropy power spectrum toward lower $\ell$ by increasing the
acoustic horizon at last scattering.  Together, the above-mentioned
effects lead to significant alterations of the CMB temperature
anisotrpy spectrum.  However, an even more pronounced effect can in
principle be seen in the CMB polarization power spectrum, which is
more sensitive to the width of the last scattering surface.
Therefore, measurements of the CMB polarization could be a critical
test for investigation of the ionization history of the Universe.

\begin{figure}
\epsfxsize=\linewidth\epsfbox{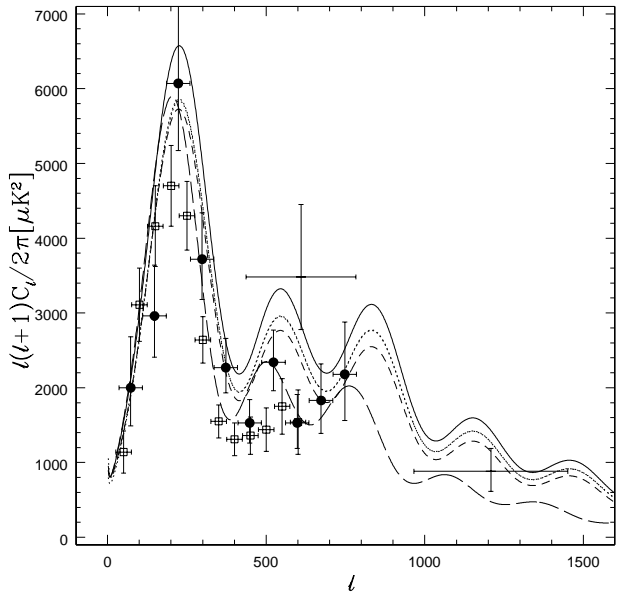}
\caption{
\label{fig:temperature}
The temperature power spectrum predicted for the various models
summarized in Table~\protect\ref{tab:chisquare}.  The lines show
predictions obtained with a modified version of CMBFAST.
The solid line corresponds to the standard model, while delayed
recombination and tilted spectrum are shown in long and short dashed,
respectively.  Both the model with late reionization and the one
including tensor modes shown in dotted lines, because the curves nearly
overlap anyway.  The various symbols represent data points of the
Maxima-1 experiment (solid circles), Boomerang (open squares), and CBI
(nothing). }
\end{figure}

There is one additional reason to discuss the possible transformations
of the CMB polarization power spectrum due to a more complicated
ionization history of the Universe at the period $z\sim 10^3$ and at
the later epochs: In {\cite{schmalzing2000:constraints}} it was
discussed how a ``conventional'' cosmological model (in particular,
with $\Omega_bh^2=0.019$ as deduced from SBBN, $\Omega_m\simeq0.3$
resulting from this and the observed baryonic fraction in clusters of
galaxies and standard recombination) can fit the Maxima-1 data
provided that the universe reionized at redshifts
$z_{\mathrm{re}}\sim15-20$ corresponding to electron scattering
optical depths $\tau\sim0.1-0.2$.  This effect is illustrated in
Figure~\ref{fig:temperature} which shows the angular power spectrum
$C_\ell$ of various models as well as the Maxima-1 data taken from
{\scite{hanany2000}} (the Boomerang and CBI data are also shown,
however we shall in this paper adopt the approach of
{\scite{schmalzing2000:constraints}} and fit to the Maxima-1 data only
-- as was discussed in {\scite{schmalzing2000:constraints}} very
similar results are obtained by fitting to the combined
Boomerang+Maxima-1 data of {\scite{jaffe2000}}).  It is seen that a
COBE-normalized,``conventional'' model with $\Omega_bh^2=0.019$,
$\Omega_m=0.3$, $\Omega_\Lambda=0.7$, $\tau=0$ and other parameters
given in table~\ref{tab:chisquare} tends to lie above the
observational data of the Maxima-1 experiment (at least for
$\ell\ga50$).  Using the offset lognormal approach of
{\scite{knox1998}} and {\scite{bond2000}} we find a $\chi^2$ of 20.2
for 10 degrees of freedom allowing this model to be rejected with more
than 95\% confidence.  The effect of reionization is to suppress the
primary anisotropies (the angular power spectrum) at $\ell\ga50$
compared to COBE scales ($\ell\simeq10$) due to electron scattering.
A model with $z_{\mathrm{re}}=15$ and all other parameters remaining
unchanged is also shown in Figure~\ref{fig:temperature}.  It is clear
that this model provides a considerably better fit to the data
($\chi^2=9.8$) - this fact was used by
{\scite{schmalzing2000:constraints}} to argue that under certain
simplifying assumptions the Maxima-1 data can be used to place a lower
limit on the redshift of reionization $z_{re}>15$ (8) at the 68\%
(95\%) confidence level.  On the other hand, a $z_{\mathrm{re}}$ much
larger than 10-15 is probably not likely, at least in the framework of
current CDM structure formation scenarios - see, e.g.,
{\scite{haiman1998}} (although reionization initially must be patchy,
their work also indicates that the approximation of a sharp transition
of the ionization state of the universe from almost neutral to fully
ionized used in our models should be a reasonable first
approximation).

Even keeping $\Omega_bh^2$ and $\Omega_m$ fixed and assuming standard
recombination there are, however, (at least) two alternative ways of
suppressing the anisotropy power spectrum $C_\ell$ at $\ell\ga50$:
either by including tensor modes (gravitational waves) or by assuming
a ``red-tilted'' initial power spectrum (and in both cases $\tau=0$).
These models are also shown in Figure~\ref{fig:temperature} - the
additional parameter of the models has been determined by requiring
that the amplitude of the angular power spectrum at the first acoustic
peak (at $\ell\sim200$) is the same as that of the model with
reionization at $z_{\mathrm{re}}=15$, described above.  The additional
parameter is given in table~\ref{tab:chisquare} for the two models
and, as can be seen from Figure~\ref{fig:temperature}, both models
result in respectable fits to the data (with $\chi^2=9.5$ and 6.8,
respectively).

\begin{figure}
\epsfxsize=\linewidth\epsfbox{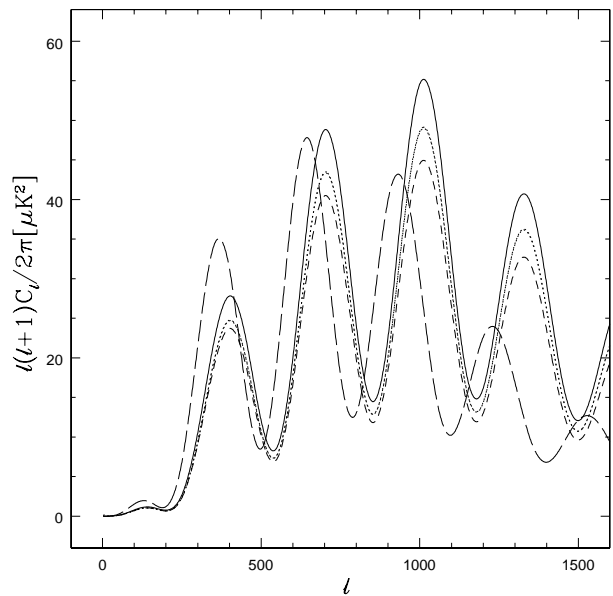}
\caption{
\label{fig:polarization}
\label{lastpage}
The polarization power spectrum.  Line styles are the same as in
Figure~\protect\ref{fig:temperature}.  }
\end{figure}

One potential way of distinguishing between the above three types of
well fitting models is by using observations of polarization
anisotropies.  The polarization power spectrum for all models
discussed above is shown in Figure~\ref{fig:polarization}.  The most
notable difference between the reionization model and the others is
the small ``bump'' at $\ell\sim5$, which is clearly seen, when the
power-spectra are plotted on a logarithmic scale. However, in the
displayed linear plot the bump is not seen and it is so small
($<1\mu$K$^2$) that it is unlikely that any of the currently planned
experiments (including the PLANCK mission) will be able to test the
existence of such a feature.

\begin{figure}
\epsfxsize=\linewidth\epsfbox{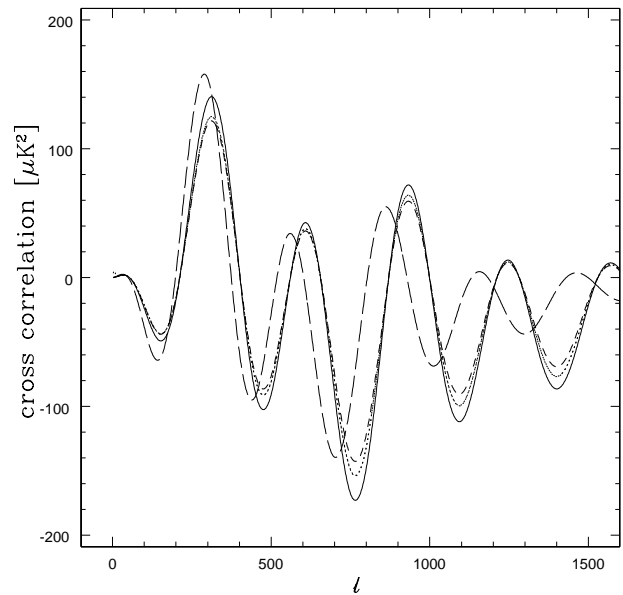}
\caption{
\label{fig:cross}
The cross correlation between temperature and polarization.  Also
here, each model is represented in the same line style as in
Figure~\protect\ref{fig:temperature}.  }
\end{figure}

We now turn to the delayed recombination models\footnote{In order to
calculate numerical spectra for the {\scite{peebles2000}} models, we
generalize the RECFAST code from {\scite{seager1999}} and the CMBFAST
code from {\scite{seljak1996:lineofsight}} to incorporate the
additional parameters.}.  We adjust the free parameter
$\varepsilon_{\alpha}$ by normalizing to the $z_{\mathrm{re}}=15$
model at the first Doppler peak, as above.  As seen from
Figure~\ref{fig:temperature}, the model provides a good fit to the
Maxima-1 data ($\chi^2=6.8$) and agrees with the recent CBI data point
published by {\scite{padin2001}} at the range $\ell\sim1000-1500$.  In
Figures \ref{fig:polarization} and \ref{fig:cross} we show the
predicted polarization power spectrum and temperature-polarization
cross correlation power spectrum for the various models on a linear
scale.  At large $\ell$ ($\sim1300-1400$) the difference between the
late recombination model and all the other models discussed becomes
considerable.  In fact, it is so large that the forthcoming PLANCK
mission is expected to be able to discriminate between the two classes
of models.  Hence, in this case polarization measurements are expected
to add valuable information to the one obtained from temperature
anisotropy measurements.

\begin{table}
\begin{center}
\begin{tabular}{lccc}
model & parameter & value & $\chi^2_{\mathrm{Maxima}}$ \\
\hline
Standard & -- & -- & 20.17 \\
Late reionization & $z_{\mathrm{re}}$ & 15 & 9.77 \\
Tensor modes & $Q_{\mathrm{T}}/Q_{\mathrm{S}}$ & 0.15 & 9.52 \\
Tilted initial spectrum & $n$ & 0.95 & 6.79 \\
Delayed recombination & $\varepsilon_\alpha$ & 7 & 6.75 \\
\end{tabular}
\end{center} 
\caption{
\label{tab:chisquare}
$\chi^2$ values for comparing the Maxima data to several variations of
a standard cosmological model.  All models share the parameters
$\Omega_bh^2=0.019$, $\Omega_\Lambda=0.7$, and $h=0.65$.  There is
neither a contribution from space curvature ($\Omega_K=0$), nor from
hot dark matter ($\Omega_\nu=0$) in any of the models.
}
\end{table}

\section*{Acknowledgement}

We are grateful to J.R. Bond, I.D. Novikov and D. Scott for
discussions.  This investigation was partly supported by INTAS under
grant number 97-1192, by RFFI under grant 17625 and by Danmarks
Grundforkskningfond through its support for TAC.

\bibliography{bibliography}

\label{lastpage}

\end{document}